\documentclass[10pt,a4paper]{article}
\usepackage[utf8]{inputenc}
\usepackage[T1]{fontenc}
\usepackage{newtxtext}
\usepackage{newtxmath}
\usepackage{amsmath}
\usepackage{amsfonts}
\usepackage{bm}
\usepackage{listings}
\usepackage{graphicx}
\usepackage[pdfa,pdfproducer={},pdfcreator={}]{hyperref}
\usepackage{url}
\usepackage{widetext}
\usepackage{mathtools}
\mathtoolsset{showonlyrefs=true}

\newcommand{\expect}[1]{\mathbb{E}\left[#1\right]}
\newcommand{\Var}[1]{\mathrm{Var}\left(#1\right)}
\newcommand{\besselI}[2]{\mathcal{I}_{#1}\left(#2\right)}
\newcommand{\besselK}[2]{\mathcal{K}_{#1}\left(#2\right)}

\usepackage{xcolor}
\usepackage{colortbl} 

\definecolor{bleudefrance}{rgb}{0.0, 0.28, 0.67}
\definecolor{bleudefrance}{rgb}{0., 0.37, 0.67}
\definecolor{resaltado}{rgb}{0.639, 0.039, 0.961}
\definecolor{histoblue}{HTML}{3981bf}
\definecolor{histobrown}{HTML}{99886e}
\definecolor{histoyellow}{HTML}{c2c099}
\definecolor{miGris}{rgb}{0.5, 0.5, 0.5}
\definecolor{verdevida}{RGB}{108,193,76}
\definecolor{verdevida2}{RGB}{56,100,39}
\hypersetup{
	colorlinks = true,
	citecolor ={bleudefrance},
	allcolors= {bleudefrance},
	linkbordercolor = {white},
}

\usepackage{tikz}
\newsavebox{\circulotachadohorizontalbox}
\sbox{\circulotachadohorizontalbox}{
	\begin{tikzpicture}[baseline=-0.5ex]
		\fill[miGris] (0,0) circle (0.5ex);
		\draw[miGris,line width=0.8pt] (-0.85ex,0) -- (0.85ex,0);
	\end{tikzpicture}
}

\usepackage[font=small,justification=RaggedRight]{caption}
\usepackage{microtype}
\usepackage[small,compact]{titlesec}
\titleformat{\section}{\bfseries\sffamily\scshape\color{black}}{\arabic{section}}{1em}{\centering\MakeUppercase}
\titleformat{\subsection}{\raggedright\bfseries\sffamily\scshape\small}{\arabic{section}.\arabic{subsection}}{1em}{\MakeUppercase}
\titleformat{\subsubsection}{\centering\bfseries\sffamily\scshape\footnotesize}{\arabic{section}.\arabic{subsection}.\arabic{subsubsection}}{1em}{\MakeUppercase}

\titleformat{\section}
{\sffamily\bfseries\scshape\normalsize}
{\arabic{section}}
{1em}
{\raggedright\MakeUppercase} 

\titleformat{\subsection}
{\sffamily\bfseries\small} 
{\arabic{section}.\arabic{subsection}}
{1em}
{\raggedright}

\titleformat{\subsubsection}
{\sffamily\itshape\small}
{\arabic{section}.\arabic{subsection}.\arabic{subsubsection}}
{1em}
{\raggedright}

\titlespacing{\subsubsection}{0pt}{*4}{*1}
\titlespacing{\subsection}{0pt}{*4}{*1}
\titlespacing{\section}{0pt}{*4}{*1}

\usepackage{fancyhdr}
\usepackage[left=1.50cm, right=1.50cm, top=1.5cm, bottom=2.00cm]{geometry}
\usepackage[backend=bibtex,isbn=true,url=true,eprint=false, bibstyle=numeric,sorting=none,    style=numeric,punctfont=true,maxbibnames=10, minbibnames=10]{biblatex}

\AtEveryBibitem{%
	\clearfield{note}%
}

\DeclareCiteCommand{\citenum}
{}
{\printfield{labelnumber}}
{}
{}

\bibliography{bibs}	

\usepackage{ragged2e}
\usepackage{booktabs}
\usepackage{tabularx}

\usepackage{placeins}

\usepackage[misc]{ifsym}

\usepackage{float}
\restylefloat{table}

\usepackage{appendix} 

\usepackage{abstract}
\usepackage{orcidlink}

\usepackage{multicol} 
\begin{document}

	\title{\Large\textsf{\textbf{{{Chromatographic Peak Shape from Stochastic Model: Analytic Time-Domain Expression in Terms of Physical Parameters and Conditions under which Heterogeneity Reduces Tailing}}}}}

	\author{\small \textsf{{Hernán R. Sánchez}}$^{1}${\scriptsize\Letter}  \orcidlink{0000-0003-1058-3396}}	
	\date{}
	
	\maketitle

	\begin{center}
	{\vspace{-0.5cm}\small 
		 Instituto de Física de Líquidos y Sistemas Biológicos,  UNLP-CONICET, La Plata, 1900, Argentina
		
		\footnotesize {\scriptsize\Letter} \textsf{\href{mailto:hernan.sanchez@quimica.unlp.edu.ar}{hernan.sanchez@quimica.unlp.edu.ar}}
		}
\end{center}

\maketitle
 \vspace{-0.3cm}

\begin{abstract}
A time-domain representation of chromatographic peak shapes is presented as an analytic expression designed for high computational efficiency, which can be used for direct time-domain peak fitting with parameters that represent physical quantities.
The underlying model integrates the effects of axial diffusion (molecular and multipath/eddy), finite initial spatial variance, and two distinct retention mechanisms: one characterized by a high rate of short-duration events (fast kinetics), and another by a low rate of long-duration events (slow kinetics).
Fits to experimental chromatograms yield substantially smaller residual standard error (RSE) than the standard EMG and the lowest average normalised RSE among 12 established peak-shape functions in the examined cases.
The stochastic approach is reformulated using single-particle probability laws, providing a rigorous basis for future theoretical extensions.  
The validity of the foundational Poisson assumption is critically examined by deriving expressions for the excess variance caused by correlated microscopic retention rate fluctuations.
A statistical interpretation of the HETP is presented and used to determine a lower bound on the number of microscopic retention events from chromatogram-derived macroscopic observables. 
This, in turn, justifies the applicability of the Gaussian limit for the mobile-plus-fast component, as established by analysis of the cumulant generating function of the closed-form benchmark derived herein. The contribution from the slow-kinetic mechanism  is incorporated via a decoupling approximation, whose validity is established through a cumulant-based analysis that explicitly bounds the decoupling-induced error. Finally, the notion that mechanistic heterogeneity necessarily exacerbates peak tailing is qualified, analytically delineating parameter regions in which it leads to a reduction in peak asymmetry.
\end{abstract}

\vspace*{0.8cm}

\begin{multicols}{2}
\setlength{\parindent}{1.0em}

\section{Introduction}
A theoretical framework that enables the prediction of chromatographic peak shapes from the parameters used to describe kinetic and transport phenomena within the column serves three key purposes: the interpretation of experimental data, the rational design of separation systems, and the development of analytical methods.
Under linear detection, the analyte-specific contribution to the chromatographic signal is proportional to the probability density function (PDF) of the elution time of any single particle of that analyte. 
In theoretical treatments, the elution time PDF is approximated by the residence-time PDF within the column. 
In the linear regime, the parameters associated with the intra-column dispersive process are concentration-independent. This property allows the overall band broadening to be treated as a superposition of independent contributions. For tractability, the temporal dispersion arising from asynchronous entries can be modeled as an initial spatial distribution determined by the injection profile. 
A rigorous route to this PDF is obtained via the stochastic approach, in which the sequence of durations of transient retention events at the stationary phase is modeled as a continuous-time stochastic process. The resulting residence-time distribution depends on the kinetic mechanisms included. The foundational Giddings-Eyring formulation\cite{EyringGiddings1955} corresponds to a minimal linear case with a single retention mechanism, and was subsequently extended to address more general conditions\cite{GiddingsCalvin1957,McQuarrie1963,Weiss1970,CavazziniRemelliDondiFrancesco1997, Cavazzini1999,felinger1999stochastic}.

Two limitations in the related literature remain salient. 
First, comprehensive mechanistic frameworks used in process design\cite{Bernau2022} are computationally demanding and do not yield analytic expressions for direct peak fitting. In parallel, many sophisticated stochastic models are formulated employing characteristic functions\cite{Dondi1986, CavazziniRemelliDondiFrancesco1997, pasti2016}, necessitating numerical inversion to obtain the time-domain peak. This alternative domain approach has limitations: it yields lower precision in parameter estimation\cite{felinger2011} and impedes a direct, intuitive interpretation of the peak shape.
 Conversely, existing time-domain treatments often do not yield a single, conveniently computable expression that couples key processes on which the retention time of each particle depends. Second, the frequent assertion that retention-site heterogeneity, a form of retention-mechanism heterogeneity, increases tailing in linear chromatography\cite{Cavazzini1999,papai2002,felinger2008} still lacks an analytically precise delineation for relevant kinetic conditions under which it holds true.

To address these issues, a rigorous and transparent stochastic–diffusive framework for linear chromatography is developed. The construction proceeds from first principles to a practical time–domain expression, with the following contributions:

(i) \textit{Foundations.} The theoretical foundations of the stochastic model are revisited and strengthened. First, an alternative derivation of the stochastic theory is presented. It elucidates the statistical origins of peak shapes by mapping the number of discrete retention events in a simplified model to specific distribution families: from Dirac delta  (no retention events) through skewed Gamma distributions (few retention events) to quasi-Gaussian profiles due to the Central Limit Theorem. The observed chromatographic peak results from a Poisson–weighted superposition of these archetypal profiles.  Second, the foundational Poisson assumption\cite{EyringGiddings1955} for the frequency of retention events  is quantified by bounding the resulting excess variance. It is shown how its suitability depends on both the magnitude and temporal correlation of the microscopic single-particle rate fluctuations, addressing a quantitative aspect previously overlooked in stochastic chromatography treatments.

(ii) \textit{Role of mechanistic heterogeneity.} The relationship between retention-mechanism heterogeneity and peak asymmetry is clarified. At a fixed expected stationary-phase time, it is shown that a system governed by two distinct retention mechanisms can, under physically relevant kinetics, exhibit lower skewness than its reference homogeneous counterpart. This demonstrates that the presence of multiple retention mechanisms does not per se imply stronger asymmetry, thus qualifying the prevailing view that retention-site heterogeneity, a specific form of retention-mechanism heterogeneity, necessarily increases tailing\cite{felinger2008}.

(iii) \textit{Link to macroscopic observables.} A statistical interpretation of the height equivalent to a theoretical plate (HETP) is established, relating it to the expected number of retention events per plate. Combined with the overdispersion analysis in (i), this yields a conservative lower bound on the total number of retention events, expressed solely in terms of chromatogram-derived quantities (plate number and retention factor). This bound is then used to justify the Gaussian approximation for the contribution from the fast-kinetic mechanism via a central-limit argument for random sums.

(iv) \textit{Time–domain peak expression.} 
A unified, analytic time-domain expression for the peak profile is derived by coupling advective–diffusive transport with the stochastic retention process. 
First, a closed-form expression for the mobile-plus-fast component is derived for an idealized Dirac delta injection and identified as a Normal Inverse Gaussian distribution, whose convergence to a Gaussian function in the high-Péclet limit is demonstrated through analysis of its cumulant generating function. 
Subsequently, the model is generalized to account for finite initial spatial variance, and the Gaussian functional form is rigorously re-established by analyzing the asymptotic limit of the underlying stochastic construction.
The practical validity of the resulting expression follows from the relative variance of the mobile-phase time, which is shown to be negligible under typical conditions. 
The expression is parameterized using the exact first two moments to ensure accuracy for finite-efficiency systems.
These parameters are linked directly to physical quantities: flow velocity, effective axial dispersion coefficient, injection spatial variance, and the retention and release rates associated with the fast-kinetic mechanism.
The contribution from an infrequent, slow-release mechanism is treated as statistically independent. The approximation is justified by demonstrating that the error introduced in the cumulants of the distribution is quantitatively negligible in the intended regime.
Consequently, the elution-time probability density is given by the convolution of the peak body with a Poisson–Gamma component corresponding to these events.
The final result is an analytic expression that incorporates confluent hypergeometric functions and allows for efficient computation. This facilitates direct peak fitting and deconvolution, circumventing the parameter estimation limitations inherent to Laplace-domain analysis.
In addition, the performance of the derived peak-shape expression is examined on two digitized asymmetric peaks from the literature, by comparing fits against all chromatographic peak-shape functions implemented in the widely used PeakFit software; in these examples, it yields the lowest average RSE among the functions tested.

\section{Theoretical Framework} 
The stochastic approach relates the chromatographic peak shape to a function proportional to the probability density function (PDF) of a random variable $T$ that describes the total residence time of a single analyte particle within the column. The analysis assumes that the detector signal originates solely from the analyte of interest, is directly proportional to its concentration, and is unaffected by extra-column dispersion. When all analyte particles enter the column synchronously and the detector response is linear, this single-particle PDF coincides—up to a constant scaling factor—with the ensemble concentration profile.  The case in which entry times are dispersed is deferred to later sections.
Focusing on the single-particle PDF facilitates an intuitive interpretation of microscopic retention processes.  
The residence time is decomposed as
$$ T = T_m + T_s, $$
where $T_m$ and $T_s$ denote the cumulative times spent in the mobile and stationary phases, respectively.  Both quantities are, in general, random variables. 

\subsection{Homogeneous model with deterministic mobile-phase residence time}
To build intuition about the influence of model parameters on peak shapes, the simplified, idealized scenario introduced above is adopted. Close resemblance to the foundational Giddings–Eyring model is retained, and a fixed mobile-phase residence time $T_m=t_m$ common to every analyte particle is explicitly imposed, with synchronous entry from the mobile phase.
Under these conditions, all variability in the total residence time $T$ arises solely from the stationary-phase component, $T_s$. 
During its passage through the column, an analyte particle may undergo multiple retention periods. Each period $i$---starting when the particle is retained by the stationary phase and ending upon its return to the mobile phase---has its residence time represented by the random variable $T_{s,i}$.

All retention events are assumed to be described by a single kinetic mechanism that remains invariant throughout the chromatographic process. The duration of each event is further assumed to be unaffected by previous events. Consequently, the random variables $T_{s,i}$ are modeled as independent and identically distributed (i.i.d.).
 Since $t_m$ is considered constant and particles enter and exit the column via the mobile phase, the specific spatial paths during stationary phase retention are irrelevant in this simplified model.

For the $T_{s,i}$'s an exponential distribution is adopted: $T_{s,i} \sim \operatorname{Exp}(\beta)$. Using this distribution is a reasonable approximation for common retention mechanisms in chromatography, such as adsorption at low analyte concentration, and applicable to any retention mechanism that can be represented by first-order kinetics, the validity of which is especially relevant at low concentration.

The total time spent in the stationary phase for a molecule experiencing $n \geq 1$ retention events is given by the random variable:
\begin{equation}
	T_{s}^{(n)} = \sum_{i=1}^{n} T_{s,i}
\end{equation}
For $n=0$ (no retention events), let $T_s^{(0)}$ denote the residence time; this variable follows a degenerate distribution, taking the value 0 with probability 1. For $n \geq 1$, it is a well-known statistical result that the sum of $n$ i.i.d. exponential random variables with rate parameter $\beta$ follows a Gamma distribution with shape parameter $n$ and rate parameter $\beta$; i.e., $T_{s}^{(n)} \sim \operatorname{Gamma}(n, \beta)$.

Insight into the peak evolution can be gained from an examination of the $n$-conditional profiles. For $n=0$, the profile collapses to a point mass at $t_m$. This component is the source of an artifact--a delta function at $t_m$--that, due to the $T_m=t_m$ idealization, inevitably appears in the final idealized peak resulting from a weighted mixture of the individual components. For $n \ge 1$, the contributions follow a sequence of $\operatorname{Gamma}(n, \beta)$ distributions shifted by $t_m$. At $n=1$, this is simply an exponential decay. As $n$ increases, these Gamma distributions grow progressively symmetric and approach a Gaussian form, consistent with the Central Limit Theorem.

\subsubsection{Distribution of the Number of Retention Events\label{sec:DistNumRetentionEvents}}
Let $f_{T_s}(t)$ denote the PDF of the total residence time in the stationary phase, $T_s$, and let $f_{T_s^{(n)}}(t)$ denote the conditional PDF given exactly $n$ retention events (i.e., the PDF of $T_s^{(n)}$). The overall PDF $f_{T_s}(t)$ is not equivalent to $f_{T_s^{(n)}}(t)$ for any single $n$, but is instead a mixture distribution. It is constructed as a sum where, for $n \ge 1$, each conditional PDF $f_{T_s^{(n)}}(t)$ is weighted by $P(N=n)$, the probability of observing exactly $n$ retention events. The sum also incorporates the term for $n=0$, which involves a Dirac delta function $\delta(t)$ at $t=0$ weighted by the corresponding probability $P(N=0)$. This relationship is formalized as:
\begin{equation}\label{sumHomoGeneric}
	f_{T_s}(t) = P(N=0)\delta(t) + \sum_{n=1}^{\infty} P(N=n) f_{T_s^{(n)}}(t).
\end{equation}

Under a constant probability per unit time of retention of an analyte particle from the mobile phase, the number of such retention events during the transit time $t_m$ can be modeled as a Poisson random variable $N \sim \operatorname{Poisson}(\Lambda)$, where  $\Lambda = \expect{N}$ is the expected number of events and $\mathbb{E}$ denotes the expectation operator. Giddings and Eyring’s seminal work relied on this Poisson model, arguing that using an average retention rate is satisfactory if the instantaneous rate fluctuates multiple times between successive retention events. In the analysis below, the conditions under which this approximation holds are examined, and the departure from Poisson statistics is quantified.

Retention dynamics subject to rate fluctuations are modeled via a stochastic process $\left\{\lambda(t): t \in [0, t_m]\right\}$. For the purpose of this specific derivation, $t$ will represent the time accrued in the mobile phase. This process is defined exclusively for a single analyte particle during its residence in the mobile phase. For each cumulative mobile‑phase residence time $t$, the non-negative random variable $\lambda(t)$ equals the hazard rate for  a retention event at $t$. Notably, $\lambda$ characterizes the microscopic retention behavior of an individual particle and is not equivalent to macroscopic kinetic constants, such as adsorption rate coefficients.

Let the interval $[0, t_m]$ be partitioned by $n+1$ points $0 = t_0 < t_1 < \dots < t_n = t_m$, defining $n$ subintervals $I_i$ where $I_i = [t_{i-1}, t_i)$ for $i=1, \dots, n-1$ and $I_n = [t_{n-1}, t_n]$. These subintervals have equal duration $\Delta t = t_m/n$. For sufficiently large $n$, $\lambda(t)$ is assumed approximately constant within each subinterval $I_i$. 
This local constancy assumption allows the number of retention events within the interval $I_i$, $N_i$, to be modeled as a conditionally Poisson-distributed random variable, $N_i \mid \lambda_i \sim \operatorname{Poisson}(\lambda_i \Delta t)$, where $\lambda_i$ is a random variable representing the effective value of  $\lambda(t)$ within $I_i$ (e.g., $\lambda_i = \lambda(t_i^*)$ for some representative $t_i^* \in I_i$). 
The expected value $\expect{N}$ can be derived via the law of total expectation. Assuming the underlying rate process is wide-sense stationary with mean $\mu_\lambda = \expect{\lambda(t)}$:
\begin{align*}
	\expect{N} &= \expect{\expect{N |\{\lambda_k\}_{k=1}^n}} = \expect{\expect{\sum_{i=1}^n N_i \bigg|\{\lambda_k\}_{k=1}^n}} \\
	&= \expect{\sum_{i=1}^n \expect{N_i | \lambda_i}} = \expect{\sum_{i=1}^n \Delta t \lambda_i}\\
	&= \sum_{i=1}^n \Delta t \expect{\lambda_i} = \sum_{i=1}^n \Delta t \mu_\lambda = t_m \mu_\lambda
\end{align*}
\noindent Thus, the expected number of retention events coincides with that of a homogeneous Poisson process with the average rate $\mu_\lambda$, even when $\lambda(t)$ fluctuates. While the mean is preserved, variability in $\lambda(t)$ affects higher-order moments; thus, matching the mean is insufficient to justify the Poisson approximation for the distribution of $N$.
A key property of the Poisson distribution is the equality of its variance and mean: $\Var{X}=\expect{X}$ for a generic Poisson variable $X$. Therefore, to assess the quality of the approximation, it is reasonable to compare the variance of $N$ in the time-varying rate model to the benchmark, $\expect{N}=t_m \mu_\lambda$. An expression for the variance of $N$ can be obtained using the law of total variance:
\begin{align}\label{eq:varConditional}
	\Var{N} &= \expect{\Var{N | \{\lambda_k\}_{k=1}^n}} + \Var{\expect{N | \{\lambda_k\}_{k=1}^n}}.
\end{align}
Given fixed rates $\lambda_i$, the counts $N_i$ in disjoint intervals are independent Poisson variables. Thus, 
\begin{equation}
	\Var{N |\{\lambda_k\}_{k=1}^n} = \sum_{i=1}^n \Var{N_i | \lambda_i} = \sum_{i=1}^n \Delta t \lambda_i.
\end{equation}
Taking the expectation yields the first term of Eq. \ref{eq:varConditional}: 
\begin{equation}\label{eVar}
	\expect{\Var{N | \{\lambda_k\}_{k=1}^n}} = \expect{\sum_{i=1}^n \Delta t \lambda_i} = t_m \mu_{\lambda}.
\end{equation}
This term is identical to the variance of a simple Poisson process with mean $t_m \mu_{\lambda}$.

The second term in Eq. \eqref{eq:varConditional} involves the variance of the conditional expectation derived earlier. Relying on the (assumed) wide-sense stationarity, $\lambda(t)$ has variance $\sigma_\lambda^2 := \operatorname{Var}(\lambda(t))=\operatorname{Var}(\lambda_i)$ and an autocorrelation function $\rho_\lambda(\tau) = \operatorname{Cov}(\lambda(t), \lambda(t+\tau)) / \sigma_\lambda^2$. The covariance between samples is approximately $\operatorname{Cov}(\lambda_i, \lambda_j) \approx \sigma_\lambda^2 \rho_\lambda(|t_i^* - t_j^*|)$. If the representative times $t_i^*$ are chosen to be regularly spaced, $\operatorname{Cov}(\lambda_i, \lambda_j) \approx \sigma_\lambda^2 \rho_\lambda(|i-j|\Delta t)$. Then,
\begin{align}
	\operatorname{Var}\left( \sum_{i=1}^n \Delta t \lambda_i\right) &= \Delta t^2 \operatorname{Var}\left( \sum_{i=1}^n \lambda_i\right) \nonumber \\ 
	&= \Delta t^2 \left( \sum_{i=1}^n \operatorname{Var}(\lambda_i) + 2 \sum_{1 \le i < j \le n} \operatorname{Cov}(\lambda_i, \lambda_j) \right) \nonumber \\
	&\approx \Delta t^2 \left[ n \sigma_\lambda^2 + 2 \sigma_\lambda^2 \sum_{1 \le i < j \le n} \rho_\lambda((j-i) \Delta t)\right] \label{eq:VarE_expanded_v3} \\ 
	&= \frac{t_m^2 \sigma_\lambda^2}{n} \left[1 + \frac{2}{n} \sum_{1 \le i < j \le n} \rho_\lambda((j-i) \Delta t)\right]. \label{eq:VarE_final_form_v3}
\end{align}

The dominant contributions to the summation arise from those terms where the time lag $(j-i)\Delta t$ is not much larger than the correlation time, $\tau_{\text{corr}}$, of $\lambda(t)$. 
Provided that $t_m \gg \tau_{\text{corr}}$, a condition met in the chromatographic process, and $n \gg \tau_{\text{corr}}/\Delta t$, a condition on the discretization that can be enforced, the sum can be approximated as:
\[
\frac{1}{n} \sum_{1 \le i < j \le n} \rho_\lambda((j-i) \Delta t) \approx \sum_{k=1}^{\infty} \rho_\lambda(k \Delta t).
\]
This approximation connects the variance calculation to the effective number of samples in $[0, t_m]$, $n_{\text{eff}}$, which can be approximated by the formula:
\begin{align*}
	n_{\text{eff}} \approx \frac{n}{1 + 2 \sum_{k=1}^\infty \rho_\lambda(k \Delta t)}, 
\end{align*}
allowing the variance of the conditional expectation to be compactly expressed as:
\begin{align*}
	\Var{\expect{N | \{\lambda_k\}_{k=1}^n}} \approx \frac{t_m^2 \sigma_\lambda^2}{n_{\text{eff}}}.
\end{align*}

Substituting the derived components for the conditional variance (Eq. \eqref{eVar}) and the variance of the conditional expectation (above) back into the law of total variance (Eq. \eqref{eq:varConditional}), the approximate total variance for the number of events $N$ is obtained:
\begin{equation}
	\Var{N} \approx \expect{N} + \frac{t_m^2 \sigma_\lambda^2}{n_{\text{eff}}} = t_m \mu_{\lambda} + \frac{t_m^2 \sigma_\lambda^2}{n_{\text{eff}}}. \label{eq:VarN_final}
\end{equation}

\noindent The first term in Eq. \eqref{eq:VarN_final} corresponds precisely to that of a Poisson process with mean $t_m \mu_\lambda$. The second term is non-negative and represents the excess variance arising from stochastic fluctuations. This result is consistent with the canonical over-dispersion in mixed Poisson processes\cite{Willmot1986} and provides an explicit analytic form of the excess variance derived from the fundamental properties of the microscopic retention rate. The validity of the Poisson approximation for the distribution of $N$ requires this excess variance to be negligible compared to the first term:
\begin{equation}
	\frac{\sigma_\lambda^2 t_m }{\mu_{\lambda} n_{\text{eff}}} \ll 1.
\end{equation}
The factor $t_m/{n_{\text{eff}}}$ represents the effective time required to obtain the statistical equivalent of one independent sample from the correlated process $\lambda(t)$, in terms of the impact on variance reduction. The effect of an increment of \( t_m \) on the relative importance of this excess variance depends on how \( t_m \) increases. If it increases due to a longer column at constant flow velocity, \( n_{\text{eff}} \) scales proportionally, $t_m/{n_{\text{eff}}}$ remains approximately constant and the relative excess variance is expected to be unaltered. However, if  $t_m$ increases due to a reduction in flow velocity at constant column length, the underlying correlations might persist for a larger fraction of the transit time, potentially increasing $t_m/{n_{\text{eff}}}$. This could therefore increase the relative contribution of the excess variance, potentially weakening the Poisson approximation at lower flow rates. 

Finally, an assessment of the validity of the Poisson approximation requires an analysis of the interdependence of the mean rate $\mu_{\lambda}$ and its variance $\sigma_\lambda^2$, since both stem from the same physical process and thus cannot be treated as independent \textit{a priori}. The validity of this approximation ultimately hinges on how $\mu_\lambda$ and $\sigma_\lambda^2$ relate within the system’s microscopic model.
To elucidate how this dependence might emerge from underlying physical constraints, consider a stylized binary model reflecting the short-range nature of common interactions leading to retentions, like those from adsorption processes: let the instantaneous rate be a constant effective rate $\lambda_0$ when the analyte is sufficiently close to the stationary phase surface, an event occurring with probability $p$, and zero otherwise.
The model yields a mean rate $\mu_\lambda = \lambda_0 p$ and variance $\sigma_\lambda^2 = \lambda_0^2 p (1-p)$. The variance-to-mean ratio simplifies to $\sigma_\lambda^2 / \mu_\lambda = \lambda_0 (1-p)$.  
In the physical limit of a narrow interaction zone ($p \ll 1$), this ratio approaches $\lambda_0$. Hence, the explicit dependence of Eq.~\eqref{eq:VarN_final}
 on $\sigma_\lambda^2/\mu_\lambda$ weakens or becomes indirect in this framework. 
Ultimately, the impact of $\sigma_\lambda^2/\mu_\lambda$ on the Poisson approximation is dictated by the underlying microscopic kinetics that couple mean retention rates with their temporal fluctuations.

\subsubsection{Residence–Time Distribution within the Column and Its Statistical Moments}\label{sec: homogeneuos model}
Assuming henceforth that the retention‑event count follows a Poisson distribution with mean $\Lambda$, and substituting the corresponding probabilities $P(N=n)$ together with the conditional densities $f_{T_s^{(n)}}(t)$ discussed in Section \ref{sec:DistNumRetentionEvents}, Eq.~\eqref{sumHomoGeneric} becomes
\begin{equation}\label{eq:fts_series}
	f_{T_s}(t)
	= e^{-\Lambda}\,\delta(t)
	+ \sum_{n=1}^{\infty}
	\frac{\Lambda^{\,n}e^{-\Lambda}}{n!}\,
	\frac{\beta^{\,n} t^{\,n-1}e^{-\beta t}}{(n-1)!},
	\qquad t\ge 0.
\end{equation}
Note that \(t\) denotes the cumulative time already spent in the stationary phase. This sum can be resummed in closed form as
\begin{equation}\label{eq:fts_closed}
	f_{T_s}(t)
	= e^{-\Lambda}\,\delta(t)
	+ e^{-\Lambda-\beta t}\,\Lambda\beta\,
	\Bigl[ \besselI{0}{2\sqrt{\Lambda\beta t}}
	- \besselI{2}{2\sqrt{\Lambda\beta t}}\Bigr],
\end{equation}
where $\besselI{\nu}{\cdot}$ denotes the modified Bessel function of the first kind and order~$\nu$. Unlike the $\mathcal{I}_1$-based expression prevalent in the literature\cite{EyringGiddings1955,McQuarrie1963,Cavazzini1999,dimarco2001}--typically stated for $t>0$ and whose rigorous extension to include the origin $(t=0)$ would require a piecewise clause--Eq.\eqref{eq:fts_closed} is a single, unified expression valid on $[0,\infty)$.
With perfectly synchronous entry of all molecules and a fixed mobile-phase time $t_m$, the single-molecule PDF $f_{T_s}(t)$, shifted by $t_m$, is directly proportional to the ensemble concentration profile generated solely by stationary-phase interactions.

It can be shown that the \(m\)-th raw moment of \(T_s\), defined as \(\mu_m := \expect{T_s^{\,m}}\), satisfies
\begin{equation}\label{eq:mu_m_homogeneous_revA}
	\mu_m \;=\; \beta^{-m}\!
	\sum_{j=1}^{m} \binom{m}{j}\,
	\frac{(m-1)!}{(j-1)!}\,
	\Lambda^{\,j}, 
	\qquad m\in\mathbb{N}.
\end{equation}
This equation is algebraically equivalent to expressions previously reported in the literature, e.g, ref. \citenum{Denizot1975}.
The first raw moment, \(\mu_1 = \Lambda/\beta\), represents the mean residence time in the stationary phase and, as anticipated, corresponds to the product of the mean number of retention events, \(\Lambda\), and the mean duration of an individual retention event, \(1/\beta\).
Let $\sigma^{2}:=\mu_2-\mu_1^{2}$ denote the variance. Substituting \(\mu_1,\mu_2\) from Eq.  \eqref{eq:mu_m_homogeneous_revA} yields  $\sigma^{2}=2\Lambda/\beta^{2}$.

To quantify the asymmetry of the \(T_s\) distribution, Fisher's moment coefficient of skewness, \(\gamma\), is employed. This coefficient is defined as:
\[
\gamma := \frac{\expect{(T_s-\mu_1)^3}}{\sigma^3} = \frac{\mu_3 - 3\mu_1\sigma^2 - \mu_1^{\,3}}{\sigma^3}.
\]
\noindent The decision to use this coefficient stems from its particular sensitivity to the tails of the distribution, a significant aspect for analyzing chromatographic tailing.  For the homogeneous model under consideration, this simplifies to 
\begin{equation}\label{eq:homoSkewness}
	\gamma = \frac{3}{\sqrt{2\Lambda}}.
\end{equation}
Thus, the skewness $\gamma$ depends solely on the mean number of retention events, $\Lambda$, whereas the rate parameter $\beta$ acts only as a time-scaling factor. As $\Lambda$ increases, the Poisson weights concentrate on large-$n$ $\Gamma(n,\beta)$ components that are nearly symmetric; the mixture $f_{T_s}(t)$ therefore inherits their quasi-Gaussian form, and its skewness decreases as $3/\sqrt{2\Lambda}$.

\section{Tailing and mechanistic heterogeneity\label{sec: tailing}}
Retention-site heterogeneity, a specific form of retention-mechanism heterogeneity, is often cited as a source of peak asymmetry.
In this analysis, that assertion is qualified by determining how the aforementioned heterogeneity affects Fisher's skewness within the canonical stochastic framework.

The skewness of a reference system, in which all retention events proceed via a single mechanism (mechanism $A$), is contrasted with that of a system where retention events are partitioned between mechanism $A$ and a kinetically distinct mechanism $B$.
To ensure an unbiased comparison, the expected stationary-phase residence time of a particle, $\expect{T_s}$, is held equal for both systems.
Under this constraint, the effect of kinetic heterogeneity is isolated by positioning both peaks at an equivalent time stage in their natural evolution towards a more symmetric, quasi-Gaussian profile.

Since the cumulative times spent in the stationary phase associated with each kinetic mechanism  are treated as independent random variables, the total stationary-phase residence time equals the sum of two independent contributions.  
Retention under each mechanism proceeds with its own constant hazard rate, and the associated residence times are independent, identically distributed, and specific to that class. The PDF of the total time in the stationary phase is therefore the convolution of the two class PDFs, allowing raw moments of the sum to be expressed as 
\[ \expect{(X+Y)^n} = \sum_{k=0}^{n} \binom{n}{k} \expect{X^k} \expect{Y^{n-k}} \]
\noindent where $\expect{X^k}$ and $\expect{Y^{n-k}}$ denote the $k$-th and $(n-k)$-th raw moments of $X$ and $Y$, respectively, and $\binom{n}{k}$ is the binomial coefficient. Throughout this section, subscripts $A$ and $B$ label quantities for  retention mechanisms $A$ and $B$, whereas the composite subscript $AB$ refers to the heterogeneous system as a whole. The relation above is used to compute the skewness shown below, whereas the mean and the variance follow directly from the additivity of first moments and second central moments. 
 \begin{align}
	\mu_{1,AB} &= \frac{\Lambda_A}{\beta_A}+ 	\frac{\Lambda_B}{\beta_B} \\\label{heteroVariance}
	\sigma_{AB}^2&=2\frac{\Lambda_A}{\beta_A^2} + 2\frac{\Lambda_B}{\beta_B^2}\\\label{heteroSkewness}
	\gamma_{AB}&= \frac{3}{\sqrt{2}}\frac{\Lambda_A \beta_B^3 + \Lambda_B \beta_A^3}{(\Lambda_A \beta_B^2 + \Lambda_B \beta_A^2)^{\frac{3}{2}}}
\end{align}

The effect of kinetic heterogeneity is assessed by comparing two chromatographic columns.  
The first, serving as a reference, is a system in which all retention events proceed via a single kinetic mechanism (mechanism $A$), yielding an average of $\Lambda_A$ retention events per analyte particle.
The second is a system where retention events also proceed via a kinetically distinct mechanism (mechanism $B$), contributing an average of $q_B\Lambda_B$ events per particle ($0<q_B\Lambda_B<\Lambda_A$). 
To preserve an identical mean retention time in both systems, the number of type-$A$ sites in the heterogeneous column is reduced to a fraction $q_A$ ($0<q_A<1$) of its original value, yielding an average of $q_A\Lambda_A$ retention events per particle on type-$A$ sites.  Consequently, conservation of the mean retention time imposes
\begin{equation}\label{eq:mean time preserved}
\frac{\Lambda_A}{\beta_A}
\;=\;
\frac{q_A\Lambda_A}{\beta_A}
\;+\;
\frac{q_B\Lambda_B}{\beta_B}.
\end{equation}
Solving this constraint for $q_{B}$ and substituting $\Lambda_{A}$ by $q_{A}\Lambda_{A}$ and $\Lambda_{B}$ by $q_{B}\Lambda_{B}$ into the expression for the skewness of the heterogeneous system ($\gamma_{A,B}$) allows a direct comparison with that of the homogeneous case ($\gamma$) via the quotient $\gamma_{A,B}/\gamma$.  Introducing the release-rate ratio $r :=\beta_{B}/\beta_{A}$, the skewness quotient can be expressed after straightforward algebra as:
\begin{equation}\label{quotientTailing}
\frac{\gamma_{A,B}}{\gamma}
\;=\;
\frac{q_{A}r^{2}-q_{A}+1}
{\sqrt{r}\,\bigl(q_{A}r-q_{A}+1\bigr)^{3/2}}.
\end{equation}

With this expression, the comparison can be reduced to two parameters, $q_A$ and $r$ (Fig. 1). For any $q_A \in (0,1)$, heterogeneity diminishes asymmetry (i.e., $\gamma_{1,2} < \gamma$) rather than invariably amplifying it, provided $r$ falls within the range:
\footnotesize
\begin{align}\label{rInec}
	1 < r &< \frac{2}{3}- \frac{2^\frac{1}{3} (-3 q_A ^3 - q_A ^4)}{3 q_A ^2 (27 q_A ^4 - 9 q_A ^5 - 2 q_A ^6 + 3 
		\sqrt{3}
		\sqrt{27 q_A ^8 - 22 q_A ^9 - 5 q_A ^{10}})^\frac{1}{3}} \\
	&+ \frac{(27 q_A ^4 - 9 q_A ^5 - 2 q_A ^6 + 3 \sqrt{3} \sqrt{27 q_A ^8 - 22 q_A ^9 - 5 q_A ^{10}})^\frac{1}{3}}{3\cdot2^\frac{1}{3} q_A ^2}
\end{align}
\normalsize

A reduction in skewness occurs only for values of $r$ greater than 1 but less than a threshold determined by $q_A$; for values exceeding this threshold, the trend reverses. The reason for this reversal becomes evident in the limit as $r \to \infty$. 
In this limit, the contribution from mechanism $B$ to the skewness in Eq.~\eqref{heteroSkewness} vanishes, and the system behaves as one in which solely a reduced number of mechanism-$A$ events ($q_A\Lambda_A$) occur.

According to Eq.~\eqref{eq:homoSkewness}, this decrease in the total event count results in a higher skewness, increased by a factor of $1/\sqrt{q_A}$. This reasoning is confirmed by the limit of the skewness ratio in Eq.~\eqref{quotientTailing}, which also converges to $1/\sqrt{q_A}$ as $r \to \infty$.

To intuitively understand this behavior, the process can be conceptualized as being equivalent to a sequence of two distinct phases: an initial phase in which only mechanism-B events occur, followed by a final phase in which only mechanism-A events occur. This abstract sequence is a generalization consistent with the simpler mental picture of a particle traversing two sequential column segments, each containing only one type of retention site. This conceptualization is justified because, under the independence assumption, the total stationary-phase residence time depends solely on the number of events associated with each mechanism, not on the sequence in which they occur. Thus, the precise spatial arrangement of sites along the column is immaterial to the resulting skewness. When $\beta_B$ is sufficiently large, the term $2q_B\Lambda_B/\beta_B^2$ in the expression of the variance becomes negligible, leaving the variance effectively determined by $2q_A\Lambda_A/\beta_A^2$. The analyte therefore reaches the segment containing sites of type $A$ with a still-narrow peak, and its final width and skewness are fixed there, where the process is equivalent to a system in which only mechanism-$A$ events occur.

\begin{center}
	\includegraphics[width=1.0\linewidth]{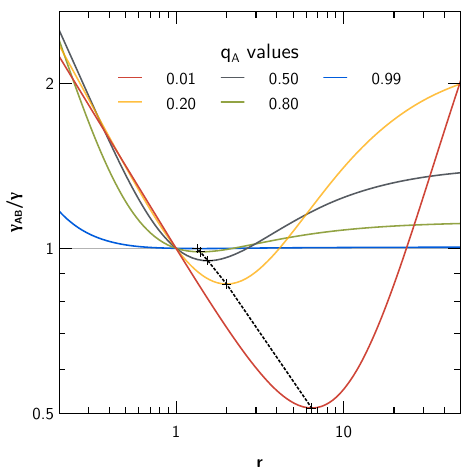}
\end{center}	
\captionof{figure}{Skewness ratio \(\gamma_{AB}/\gamma\) versus release-rate ratio \(r=\beta_B/\beta_A\) for selected fractions \(q_A\). Values below unity indicate heterogeneity-induced reduction of tailing.}
\label{fig:gammaquotientvsr}

\setlength{\parindent}{1.0em}

$\,$ \\

When $q_A$ is very small, the retention process is almost exclusively determined by the statistics of mechanism-$B$ events, approaching the behavior of a system with a single retention mechanism. A reduction in tailing is achieved by increasing the mean number of mechanism-$B$ retention events, $q_B\Lambda_B$. 
Because the mean stationary-phase time is fixed, this increase implies a proportional increase in the release-rate $\beta_B$. Consequently, the skewness, which is approximately $\gamma \approx 3/\sqrt{2q_B\Lambda_B}$, can in principle be arbitrarily reduced towards zero.

For a given $q_A$, the $r$ value that minimizes $\gamma_{1,2}/\gamma$ is  
\[
r^*(q_A)=\frac{2}{3}+\frac{1}{3}\sqrt{1+\frac{3}{q_A}}.
\]  
Substituting this into Eq.~(\ref{quotientTailing}) yields  
\[
\frac{\gamma_{1,2}(q_A)}{\gamma} =
\frac{4}{\sqrt{2+\sqrt{1+3/q_A}}\,
	\sqrt{\sqrt{q_A^{2}+3q_A}+3-q_A}}.
\]

\section{Time–domain peak expression obtained by coupling axial diffusion with stochastic retention} \label{sec:diffusion}

In this section, an analytical time-domain treatment of chromatographic peak shape is presented. It couples the contributions from advective-diffusive transport (including axial dispersion and finite initial spatial variance) with a stochastic retention process comprising two distinct kinetic mechanisms: one fast and one slow. The model relies on physically justified approximations that are highly accurate for typical chromatographic conditions. The resulting expression is computationally efficient, offering a significant advantage over methods requiring numerical inversion from the Laplace domain for the direct analysis of experimental chromatograms.

The total residence time, $T$, of an analyte particle within the column is modeled as the sum of three distinct random variables, each associated with a fundamental physical process. First, $T_m$ is the first-passage time required for the particle to traverse the column of length $x$ while residing exclusively in the mobile phase. Its transport is due to advection, with mean velocity $v$, and longitudinal diffusion, with coefficient $D$. The probability density function (PDF) for this process is the well-established inverse-Gaussian distribution $\mathrm{IG}\!\left(\mu=\frac{x}{v},\;\lambda=\frac{x^2}{2D}\right)$ with the conventional parametrisation used here:
\begin{equation}
	f_{T_m}(t_m) = \frac{x}{2\sqrt{\pi D t_m^3}} \exp\left[-\frac{(x-vt_m)^2}{4Dt_m}\right]
\end{equation}
where $t_m$ is a realization of the random time $T_m$.

Second, the cumulative times spent in the stationary phase that are associated with the fast (subscript $f$) and slow (subscript $s$) kinetic mechanisms, $T_{s,f}$ and $T_{s,s}$, are treated as conditionally independent given $T_m=t_m$. Following the model developed in previous sections, their respective conditional distributions will be treated as compound Poisson-Gamma model (Eqs. \eqref{eq:fts_series} and  \eqref{eq:fts_closed}). The model is parameterized by expressing the mean number of retention events for each mechanism $i\in \{f,s\}$ as $\lambda_i t_m$, where $\lambda_i$ represents the constant retention rate for mechanism $i$.

Obtaining an analytic-form for the overall PDF appears intractable using standard analytical approaches; the component processes are statistically dependent and already involve complex expressions. Although the series representation (Eq. \ref{eq:fts_series}) has simpler per-term factors than the Bessel-functions form, it is impractical for the fast component $T_{s,f}$. This is because $\lambda_f t_m$ is typically large, accurate evaluation requires many non-negligible terms, which is computationally impractical. 
Consequently, a convenient strategy to find an analytic solution is to employ highly accurate approximations, justified by robust physical and mathematical arguments, as detailed below.

\subsection{A Statistical Foundation for the Height Equivalent to one Theoretical Plate}
An interpretation of the height equivalent to a theoretical plate, \(H\), is developed.
This is relevant for two reasons.
First, it provides an intuitive view of this key chromatographic metric by relating macroscopic operational parameters to microscopic events. In particular, it relates macroscopic efficiency measures (plate number and retention factor) to the expected number of microscopic retention events. 
Second, this connection establishes a lower bound on the event count, which is used in the next section to justify the applicability of the Gaussian limit for the mobile-plus-fast component.

In typical HPLC columns, a segment of length $H$ spans many characteristic microstructural units of the bed (e.g., several packing particle diameters or comparable structural lengths), i.e., a mesoscopic scale much larger than microscopic relaxation lengths.
Under this scale separation, and assuming axial homogeneity at this scale along the column, the distribution of the per-segment residence time for segments of length $H$ is (approximately) position-invariant (spatial stationarity), and inter-segment correlations are short-ranged. 
This justifies the use of constant per-unit-length rates, $\varepsilon$ and $\nu$, as defined below:
\[
\varepsilon:=\frac{\operatorname{E}[T_H]}{H}, \qquad \nu:=\frac{\operatorname{Var}(T_H)}{H},
\]
where $T_H$ is the random variable representing the residence time in a single segment of length $H$. By linearity of the expectation operator, the total mean residence time of a single particle in a column of length $L$, denoted $T(L)$, is:
\[
\expect{T(L)} = \varepsilon L = \frac{L}{H}\expect{T_H}
\]
Since the microscopic correlation length is much smaller than the segment length $H$, the covariance between the residence times of any two non-overlapping segments is negligible. This short-range dependence justifies the linear scaling of the total variance:
$$\Var{T(L)} =:\sigma_T^2(L)\approx \nu L = \frac{L}{H}\Var{T_H}.$$
Because $\expect{T}$ is often a very good approximation to the reported retention time \(t_R\) (exact for symmetric-unimodal peaks in the absence of extra-column effects), the plate-number ($N'$) definition yields:
\[
N' = \frac{t_R^2}{\sigma_T^2(L)} \approx \frac{(\expect{T(L)})^2}{\Var{T(L)}} = \frac{L}{H} \frac{(\expect{T_H})^2}{\Var{T_H}}
\]
Under these mild approximations, the length $H$ equals the theoretical plate height ($L/N'$) if 
\[\expect{T_H}^2 = \Var{T_H}.\]
This result offers a direct statistical interpretation of the Height Equivalent to a Theoretical Plate. It can be understood as approximately the characteristic length, $H$, of a conceptual column segment for which the variance of the residence time equals the square of its expectation.  This statistical interpretation is valid under linear, isocratic, and isothermal conditions, with axially stationary per-segment statistics, negligible extra-column effects, and non-pathological peak shapes. Consequently, it should be regarded as a useful approximation rather than a general identity.

\subsubsection{Lower Bound for the Number of Retention Events from the Chromatogram}
Let the total residence time in a segment of length $H$ be the sum $T_H = T_{m,H} + S_H$, where $T_{m,H}$ and $S_H$ are random variables representing the mobile- and stationary-phase residence times, respectively. The stationary-phase time, $S_H$, is the sum of a random number of individual retention durations, $n_H$, such that $S_H = \sum_{j=1}^{n_H} Y_j$. Conditional on $T_{m,H}$, the arrivals of these retention events are modeled as a stationary counting process with a mean rate $r$ per unit of mobile-phase time, yielding a conditional expected number of retention events $\expect{n_H \mid T_{m,H}} = r T_{m,H}$. Each duration $Y_j$ is an independent and identically distributed random variable with mean $\mu_Y$ and variance $\sigma_Y^2$.
The quantity $k' := \expect{T_s}/\expect{T_m}=r\,\mu_Y$ represents a microscopic retention factor that, under isocratic and isothermal operation, 
coincides with the operational ratio \((t_R-t_M)/t_M\) when the reported times are approximated as expectations, which is exact for symmetric unimodal peaks, and typically very accurate in practice.

By conditioning on \(T_{m,H}\) and applying the laws of total expectation and total variance:
\begin{align}
	\expect{T_H\mid T_{m,H}}
	&= T_{m,H}+\expect{S_H\mid T_{m,H}}\\
	&= T_{m,H}+k'\,T_{m,H}
	= (1+k')\,T_{m,H}, 
\end{align}
\begin{align}
	\expect{T_H}
	&= \expect{\expect{T_H\mid T_{m,H}}}
	= (1+k')\,\expect{T_{m,H}}. \label{eq:ETH-deriv}
\end{align}
And
\begin{align}
	\Var{T_H\mid T_{m,H}}
	&= \Var{S_H\mid T_{m,H}}, 
\end{align}
\begin{align}
	\Var{T_H}
	&= \Var{\expect{T_H\mid T_{m,H}}}
	+ \expect{\Var{T_H\mid T_{m,H}}} \nonumber\\
	&= (1+k')^{2}\Var{T_{m,H}}
	+\expect{\Var{S_H\mid T_{m,H}}}. \label{eq:VarTH-deriv}
\end{align}

Since $S_H=\sum_{j=1}^{N_H}Y_j$ with $Y_j\stackrel{\mathrm{i.i.d.}}{\sim}Y$, its variance stems from two sources of randomness: the number of events, $N$, and the duration of each event, $Y_j$. Using the conditional law of total variance,
\begin{align}\label{eq:cond-total-var}
	\Var{S_H\mid T_{m,H}}
	&= \expect{\Var{S_H \mid N_H, T_{m,H}) \,\big|\, T_{m,H}}}\\\nonumber
	&+ \operatorname{Var}\!\left(\operatorname{E}[S_H \mid N_H, T_{m,H}] \,\big|\, T_{m,H}\right). 
\end{align}
Thus, the inner conditional moments are
\begin{align}
	\expect{S_H \mid N_H, T_{m,H}}&=N_H\,\mu_Y\\
	\Var{S_H \mid N_H, T_{m,H}}&=N_H\,\sigma_Y^2.
\end{align}
Substituting these into \eqref{eq:cond-total-var} yields
\begin{align}
	\Var{S_H\mid T_{m,H}}
	&= \expect{N_H\,\sigma_Y^2 \,\big|\, T_{m,H}}
	\;+\; \Var{N_H\,\mu_Y \,\big|\, T_{m,H}} \nonumber\\
	&= \sigma_Y^{2}\,\expect{N_H\mid T_{m,H}}
	\;+\; \mu_Y^{2}\,\Var{N_H\mid T_{m,H}}. \label{eq:VarSH-final}
\end{align}

It was shown in Section \ref{sec:DistNumRetentionEvents} that the presence of rate fluctuations, which are expected within the segment, makes the counting mechanism over–dispersed relative to Poisson. Therefore,  \(\Var{N_H\mid T_{m,H}} \ge \expect{N_H\mid T_{m,H}}=r\,T_{m,H}\). 
By taking the expectation and employing the identity $\expect{Y^{2}}=\mu_Y^{2}+\sigma_Y^{2}$, the inequality becomes:
\begin{equation}
	\Var{T_H}\;\ge\;(1+k')^{2}\Var{T_{m,H}}
	\;+\; r\,\expect{Y^{2}}\,\expect{T_{m,H}},
	\label{eq:varTH-lb}
\end{equation}
and the inequality becomes an equality in the case of a homogeneous Poisson arrival process.

Applying the statistical plate condition previously derived, \(\operatorname{Var}(T_H)=\big(\expect{T_H}\big)^{2}\), and substituting from Eqs. \eqref{eq:ETH-deriv} and \eqref{eq:varTH-lb} yields:
\begin{equation}
	(1+k')^{2}\big(\expect{T_{m,H}}\big)^{2}\;\ge\; r\,\expect{Y^{2}}\,\expect{T_{m,H}},
\end{equation}
because \(\Var{T_{m,H}}\ge 0\). Let $\Lambda_H:=r\,\expect{T_{m,H}}$ be the expected number of retention in a segment of length $H$. Using the coefficient of variation $\operatorname{CV}_Y:=\sigma_Y/\mu_Y$ and the definition of $k^\prime$, the following conservative bound is obtained
\begin{equation}
	\Lambda_H \;\ge\; \frac{r^{2}\, \expect{Y^{2}}}{(1+k')^{2}}
	\;=\; \frac{k'^{2}\bigl(1+\operatorname{CV}_Y^{2}\bigr)}{(1+k')^{2}}.
	\label{eq:lambda-plate-lb}
\end{equation}
This inequality is tight for homogeneous Poisson arrivals and becomes stricter under any over–dispersion induced by intra–segment rate fluctuations.

At the column scale, the expected total number of retention events ($\expect{N_{\mathrm{ret,tot}}}$) is the sum of the mean counts over all segments and equals $N_H \Lambda_H$. As established above, the number of statistical stages, $N_H$, is well-approximated by the experimental plate number, $N^\prime$. Consequently, a direct link between the microscopic retention events count and the macroscopic column efficiency is established:
\begin{equation}
	\expect{N_{\mathrm{ret,tot}}} \;\ge\; \frac{k'^{2}\bigl(1+\operatorname{CV}_Y^{2}\bigr)}{(1+k')^{2}} N^\prime
\end{equation}

For the homogeneous model adopted in this work, with
$Y \sim \operatorname{Exp}(\beta)$, (thus, $CV_Y^2 = 1$ and $k'=r/\beta$), the per-plate bound \eqref{eq:lambda-plate-lb} becomes:
\begin{equation}
	\Lambda_H \ge \frac{2k'^2}{(1+k')^2}
\end{equation}
and therefore
\begin{equation}
	\expect{N_{\mathrm{ret,tot}}} \ge \frac{2k'^2}{(1+k')^2}\,N^\prime
\end{equation}
Thus, within the proposed stochastic retention model, typical values of the macroscopic observables imply a large expected microscopic event count.

\subsection{Composite Distribution for Mobile Phase and Fast Retention}
The condition established in the preceding section, that the expected fast-site event count is large, justifies the application of a central limit theorem for random sums (the version applicable when the number of terms is itself random; see, e.g. ref. \citenum{gut2012anscombe}). It follows that, conditional on \(t_m\), the cumulative fast-kinetics residence time is well approximated by a Gaussian function,
\[
T_{s,f}\mid t_m \;\approx\; \mathcal N\!\big(\mu_f t_m,\;\sigma_f^2 t_m\big)
\]
with $\mu_f=\lambda_f/\beta_f$ and $\sigma_f^2=2\lambda_f/\beta_f^2$, which is the surrogate used in what follows. These parameters follow either from the laws of total expectation and variance or, equivalently, by adapting the moment formula in Eq.~\eqref{eq:mu_m_homogeneous_revA} with $\Lambda = \lambda_f t_m$ and $\beta = \beta_f$. While the Gaussian surrogate permits non-physical values ($T_{s,f}<0$), the probability associated with this set is in practice vanishingly small. For analytical convenience, the untruncated form is therefore adopted.

Define \(f_{T_G}(t_G)\) as the probability density of the composite time \(T_G:=T_m+T_{s,f}\) at real time \(t_G>0\). For each \(t_m\), set \(t_{s,f}:=t_G-t_m\). Since \(f_{T_m}\) is undefined at \(0\), its continuous extension at the origin is adopted, \(f_{T_m}(0):=0\).
Causality is enforced by the upper limit \(t_m\le t_G\); the following integral on \([0,t_G]\) is therefore well defined:
\[
f_{T_G}(t_G)=\int_{0}^{t_G} f_{T_m}(t_m)\,f_{T_{s,f}}(t_G-t_m\mid t_m)\,dt_m.
\]
With the untruncated conditional gaussian for \(T_{s,f}\) already adopted. The upper limit can be extended to \(\infty\), as the portion \(t_m>t_G\) corresponds to \(t_{s,f}<0\) and contributes only negligible probability. With this approximation, completing the square in the exponents and grouping $t_G$-independent factors yields an integral which matches the identity 3.471.9 form ref. \citenum{gradshteyn2007}. The result is the closed form
\begin{equation}\label{eq:DistrMobileFast}
	f_{T_G}(t_G) = \frac{x\;e^{\frac{vx}{2D}+\frac{1+\mu_f}{\sigma_f^2}t_G}}{\pi\sqrt{2\sigma_f^2D}}
	\sqrt{\frac{B}{A(t_G)}}\besselK{-1}{2\sqrt{A(t_G)B}}
\end{equation}
where $\mathcal{K}_\nu$ denotes the modified Bessel function of the second kind and order $\nu$, and
\begin{align}
	A(t_G):=&	\frac{x^2}{4D}+\frac{t_G^2}{2\sigma^2_f}   \\ 
	B:=& \frac{v^2}{4D}+\frac{(1+\mu_f)^2}{2\sigma_f^2}
\end{align}
A quantitatively excellent approximation of Eq. \eqref{eq:DistrMobileFast} is obtained by applying the large-argument asymptotic expansion of $\mathcal{K}_\nu$ 
for fixed order (DLMF \S10.40.2, ref.~\citenum{NIST:DLMF}) whose leading term is
\begin{equation}
	\besselK{\nu}{z}\sim \left(\frac{\pi}{2z}\right)^\frac{1}{2}e^{-z}
\end{equation} 
Error bounds in DLMF \S10.40.11 ensure rapid convergence, so the leading term is accurate
for moderately large \(z\).
Since the argument \(z=2\sqrt{A(t_G)B}\) greatly exceeds this threshold within the parameter range of interest, the asymptotic expansion provides a quantitatively excellent approximation. Implementing this replacement gives
 \begin{equation}
	f_{T_G}(t_G)\approx \tilde{f}_{T_G}(t_G):=  \frac{xe^{\frac{vx}{2D}}B^{1/4} }{\sqrt{8\pi\sigma_f^2D}}
	\;
	\frac{e^{\frac{1+\mu_f}{\sigma_f^2}t_G}e^{-2\sqrt{A(t_G)B}}}{A(t_G)^{3/4}}
\end{equation}

The PDF in Eq. \eqref{eq:DistrMobileFast} is identical to that of a Normal Inverse Gaussian (NIG) distribution\cite{Nielsen1977} $(\mathcal{NIG}(\alpha,\beta,\delta,\mu))$:
\begin{equation}
f(t)=\frac{\alpha\delta \besselK{1}{\alpha\sqrt{\delta^2 + (t - \mu)^2}}}{\pi \sqrt{\delta^2 + (t - \mu)^2}} \; e^{\delta \gamma + \beta (t - \mu)},\,
\gamma:=\sqrt{\alpha^{2}-\beta^{2}}
\end{equation}
under the mapping
\begin{equation}
\alpha=\sqrt{\frac{v^2}{2D\sigma_f^2}+\beta^2},\quad
\beta=\frac{1+\mu_f}{\sigma_f^{2}},\quad
\delta=\frac{\sigma_f x}{\sqrt{2D}},\quad
\mu=0.
\end{equation}

It is a standard result that the NIG distribution converges to a normal distribution when $\alpha\to\infty$, $\delta/\alpha$ remains constant, and $\beta=0$\cite{Nielsen1997}. The last condition is not satisfied in the present case but can be relaxed. The cumulant generating function (CGF) for $\mathcal{N}(\mu_\mathcal{N},\sigma_\mathcal{N}^2)$ is $K_\mathcal{N}(u)=\mu_\mathcal{N} u + \sigma_\mathcal{N}^2/2 u^2$.
The CGF of the NIG is
\begin{equation}\label{eq:CGFNIG}
K_\mathcal{NIG}(u)=\mu u + \delta (\sqrt{\alpha^{2}-\beta^{2}} - \sqrt{\alpha^2 -(\beta +u)^2})
\end{equation}
Factoring $\alpha$ from the square–roots and expanding $\sqrt{1-z}$ for $|z|<1$ yields
\begin{equation}\label{eq:KNIG_approx}
K_{\mathcal{NIG}}(u)=\mu u+\frac{\delta}{\alpha}\beta u+\frac{\delta}{2\alpha}u^2
+ \dots
\end{equation}
where the omitted terms tend to zero if $\alpha\to\infty$, $\delta/\alpha$ is a finite constant and $\beta/\alpha\to0$.
This set of more general conditions provides a suitable approximation for the systems modeled herein, as standard chromatographic conditions are typically characterized by a high Péclet number ($\text{Pe}=xv/D\gg1$).
 Consequently, the remainder terms in the CGF expansion from Eq. \eqref{eq:KNIG_approx}, which represent all higher-order cumulants, are rendered numerically negligible relative to the first two cumulants. The CGF of the NIG is thus well-approximated by the Gaussian CGF. 
The Taylor expansion used to derive the limit is valid  $|(\beta+u)/\alpha|<1$, a condition that defines a valid domain for $u$ as long as $\alpha$ is finite. However, the convergence of a CGF in any open interval containing the origin is sufficient to establish convergence in distribution\cite{curtiss1942note}.
Therefore, the NIG distribution can be approximated by the limiting distribution $\mathcal{N}(\mu_G,\sigma_G^2)$
under the experimentally relevant conditions of high Péclet number. The parameters $\mu_G$ and $\sigma_G^2$ are the exact first and second cumulants of the NIG. These can be readily obtained from Eq. \eqref{eq:CGFNIG} and are known to be\cite{Nielsen1997}:
\begin{equation}
	\kappa_1 = \mu + \frac{\delta\beta}{\gamma}, \qquad
	\kappa_2 = \frac{\delta\alpha^2}{\gamma^3}.
\end{equation}
Applying the parameter mapping to these general formulas yields the exact moments for the $T_G$ distribution:
\begin{align}
	\mu_G = \kappa_1 &= (1+\mu_f)\frac{x}{v} \\
	\sigma_G^2 = \kappa_2 &=  (1+\mu_f)^2 \frac{2Dx}{v^3} +\frac{\sigma_f^2 x}{v}
\end{align}
Under typical chromatographic regimes (high Péclet number, long columns), departures between $f_{T_G}(t_G)$ and its normal counterpart are not discernible within plotting resolution, except under extreme magnification.

\subsection{Normal Approximation and non-spike injection}\label{sec:Normal Approximation}

For subsequent analytical tractability, the distribution of $T_G$ is approximated employing a normal surrogate, $\mathcal{N}(\mu_G, \sigma_G^2)$.
The parameters are chosen to match the first two moments of the distribution of $T_G$: $\mu_G := \expect{T_G}$ and $\sigma_G^2 := \Var{T_G}$. By the laws of total expectation and total variance, the following identities are obtained:
\begin{align}\label{eq:mu_Ggeneral}
	\mu_G &= (1+\mu_f)\expect{T_m}\\\label{eq:Var_Ggeneral}
	\sigma_G^2& =(1+\mu_f)^2\,\Var{T_m}+\sigma_f^2\expect{T_m}
\end{align}

When the distribution of initial particle axial-positions is modeled by a Dirac-delta function, representing an identical injection time and axial position for all molecules, the per-molecule mobile-phase residence time, $T_m$, is described by an inverse Gaussian distribution with parameters that are invariant across molecules. 

A more general result for injections with finite spatial variance is provided below. 
The injection profile is characterized as a distribution for the per-particle initial axial position, represented by the random variable $X_0$. Without loss of generality, the origin is chosen so that $x_0:=\expect{X_0}=0$. The spatial variance ($\Var{X_0}$) is denoted by $\sigma^2_0$.
In the standard regime where $x \gg \sigma_0$, the probability mass contained within the mathematically problematic domain $X_0 \ge x$ is insignificant; consequently, the effect of the truncation at $x$ on the moments of the distribution is negligible, even when $X_0$ has unbounded support.
Since $T_m$ is defined conditionally on $X_0$, the law of total expectation is applied as follows.
\begin{equation}\label{eq: ExpectTm}
	\expect{T_m}=\expect{\expect{T_m|X_0}}=\expect{\frac{x-X_0}{v}}=\frac{x}{v}
\end{equation}
Furthermore, under the same $x\gg \sigma_{0}$
assumption, the law of total variance gives
\begin{align}
	\Var{T_m}&=\expect{\frac{2D(x-X_0)}{v^3}} +  \Var{\frac{x-X_0}{v}}\\
	&= \frac{2Dx}{v^3} + \frac{\sigma_0^2}{v^2}\label{eq: VarTm}
\end{align}
The first term is due to molecular axial dispersion. If multipath (Eddy) effects are to be included, an effective longitudinal dispersion coefficient $D_{\mathrm{eff}}$ may be adopted in place of $D$, preserving the same structure. This generalized description is used for the final $\sigma_G^2$ expression. The second term maps the spatial injection variance into the time domain. Finally, substituting the above moments into Eqs. \eqref{eq:mu_Ggeneral} and \eqref{eq:Var_Ggeneral}  yields
\begin{align}
	\mu_G &= (1+\mu_f)\frac{x}{v}\\
	\sigma_G^2 &=(1+\mu_f)^2\left(\frac{2D_{\text{eff}}\;x}{v^3} + \frac{\sigma_0^2}{v^2}\right)+\sigma_f^2\frac{x}{v}.
\end{align}
As expected, the general expression reduces to the result of an idealized (zero-variance) injection, as presented in the previous section, by setting $\sigma_0=0$.

Adopting a Gaussian functional form for this more general case can be justified as follows. An alternative representation for $T_G$ is:
\begin{equation}\label{eq:TG-construction}
	T_G \;=\; (1+\mu_f)\,T_m \;+\; \sqrt{\sigma_f^2\,T_m} \, \xi
\end{equation}
where 	$\xi\sim\mathcal N(0,1)$ is independent of $T_m$. It follows from the conditional law \(T_{s,f}\mid T_m\sim\mathcal{N}(\mu_f T_m,\sigma_f^2 T_m)\) and $T_G=T_m+T_{s,f}$, as any $Y\sim \mathcal N(\mu,\sigma^2)$ can be written as $\mu + \sigma \xi$.

Based on this representation, $T_G$ can be treated as a function of two independent random variables, $T_G = g(T_m, \xi)$, where $g$ is the deterministic function:
\[
g(a,b):=(1+\mu_f)a+\sqrt{\sigma_f^2 a}\,b.
\]
Along the asymptotic trajectory approaching the high-efficiency limit, defined by $D_{\mathrm{eff}}\to 0$ and $\sigma_0/x\to 0$ for fixed $x$ and $v$, the mobile-phase time $T_m$ converges in probability to its expectation $C:=x/v$, while $\xi$ trivially converges in distribution to $\mathcal N(0,1)$. Together with the continuity of $g$, this implies, by Slutsky's\cite{casella2024statistical} and continuous-mapping\cite{van2000asymptotic} theorems, that:
\[
T_G = g\!\big(T_m, \xi\big) \xrightarrow{d}  g(C,\xi) \;=\; (1+\mu_f)C+\sqrt{\sigma_f^2 C}\,\xi.
\]
Accordingly, the Gaussian functional form arises as the exact limit by construction.

For a system with large but finite efficiency, this approach is validated by the squared coefficient of variation of $T_m$,
\[
\mathrm{CV}_{T_m}^2=\frac{\Var{T_m}}{\expect{T_m}^2}
=\frac{2}{\text{Pe}} + \left(\frac{\sigma_0}{x}\right)^2
\]
which is negligible under standard chromatographic conditions ($x\gg\sigma_0$, $\text{Pe}\gg1$). 
Thus, using the Gaussian functional form and setting ($\mu_G, \sigma_G^2$) to the finite-efficiency moments is justified on firm theoretical grounds.

\subsection{Treatment of Infrequent and Slow-Release Events}
Within the Gaussian approximation for $T_G$, the slow-mechanism contribution $T_{s,s}$ is the sole source of tailing. 
This contribution is statistically coupled to the peak body because its event count $N_s\!\sim\!\mathrm{Pois}(\lambda_s T_m)$ depends on the random mobile-phase time $T_m$. 
To obtain a final analytic solution, the processes are decoupled by replacing $T_m$ only within the Poisson parameter by its expectation $\expect{T_m}=x/v$. 
An independent surrogate $T'_{s,s}$ with fixed Poisson parameter $\Lambda_s=\lambda_s x/v$  is thus defined. 
The total time is modeled as the sum $T = T_G + T'_{s,s}$. Since $T_G$ and $T'_{s,s}$ are independent, the PDF of $T$ equals the convolution of their individual PDFs. 

The validity of this approximation is assessed by analyzing the resulting error in the cumulants
For $T_{s,s}$, which represents the sum of $N_s$ independent and identically distributed random variables $Y \sim \mathrm{Exp}(\beta_s)$, the conditional moment-generating function (MGF) is: 
$$\expect{e^{uT_{s,s}} \mid T_m, N_s} = \left(\expect{e^{uY}}\right)^{N_s}=
\left(\frac{\beta_s}{\beta_s - u}\right)^{N_s},
\, \text{for} |u| < \beta_s
$$
Given $N_s(T_m) \mid T_m \sim \text{Pois}(\lambda_s T_m)$, taking the conditional expectation with respect to $N_s$ yields:
\begin{align}
	\expect{e^{uT_{s,s}} \mid T_m} &= \sum_{n=0}^\infty \left(\frac{\beta_s}{\beta_s - u}\right)^n \frac{e^{-\lambda_s T_m}(\lambda_s T_m)^n}{n!}\\
	&=\exp\left[\lambda_s T_m \left(\frac{\beta_s}{\beta_s - u} - 1\right)\right]=\exp\left[T_m c(u)\right]
\end{align}
where 
$$c(u) := \lambda_s \left(\frac{\beta_s}{\beta_s - u} - 1\right) = \lambda_s \sum_{m=1}^\infty \frac{u^m}{\beta_s^m}$$
Consequently, the conditional CGF is 
$$K_{T_{s,s} \mid T_m}(u) = \log \expect{e^{uT_{s,s}} \mid T_m}=T_m \cdot c(u)$$
The unconditional CGF is obtained by taking the expectation with respect to $T_m$:
$$K_{T_{s,s}}(u) = \log \expect{e^{uT_{s,s}}}=\log \expect{e^{c(u)T_m }}$$
This expression corresponds to the CGF of $T_m$ evaluated at $c(u)$. Therefore, by function composition:
\begin{equation}\label{eq:cumulantTss}
	K_{T_{s,s}}(u) = K_{T_m}(c(u)) = \sum_{k=1}^\infty \frac{\kappa_k(T_m)}{k!} c(u)^k.
\end{equation}
where $\kappa_k(T_m)$ denotes the $k$-th cumulant of $T_m$.

The decoupled CGF, $K_{T'_{s,s}}(u)$, is obtained from substitution of $T_m$ by $\expect{T_m}$. This operation nullifies all cumulants $\kappa_k(T_m)$ for $k \ge 2$ in Eq. \eqref{eq:cumulantTss}, as the higher-order cumulants of a constant value ($\expect{T_m}$) are zero. Thus:
$$K_{T'_{s,s}}(u) = \kappa_1(T_m) c(u) = \expect{T_m} c(u)$$
The general $j$-th cumulant of this decoupled process is obtained from its CGF derivatives at $u=0$, yielding:
$$\kappa_j^{\text{(dec)}} = \lambda_s \expect{T_m} \cdot \frac{j!}{\beta_s^j}$$
With $\Lambda = \lambda_s \expect{T_m}$, the resulting mean, variance, and skewness coincide with those derived for the homogeneous model (Sec. \ref{sec: homogeneuos model}), thereby confirming framework consistency.

The decoupled CGF $K_{T'_{s,s}}(u)$ is exact for the mean (the $k=1$ term of Eq. \eqref{eq:cumulantTss}), a result that can also be proved via the law of total expectation. The error in all higher-order cumulants ($j \ge 2$) arises from the $k \ge 2$ terms of the expansion:
$$\kappa_j - \kappa_j^{\text{(dec)}} = \sum_{k=2}^j \frac{\kappa_k(T_m)}{k!} \left.\frac{d^j}{du^j} c(u)^k \right|_{u=0}$$

To evaluate the $j$-th derivative,  $c(u)^k$ is first expressed in a form suitable for series expansion:
$$c(u)^k = \frac{\lambda_s^k}{\beta_s^k} u^k \left( 1 - \frac{u}{\beta_s} \right)^{-k}$$
Application of the negative binomial series to the $(1 - u/\beta_s)^{-k}$ factor yields:
$$c(u)^k = \lambda_s^k \sum_{\nu=0}^\infty \binom{k+\nu-1}{\nu} \frac{u^{k+\nu}}{\beta_s^{k+\nu}}$$
The $j$-th derivative at $u=0$ equals $j!$ times the coefficient of the $u^j$ term. The latter is obtained by setting the exponent $k+\nu = j$, and replacing  $\nu$ by  $j-k$. Applying the binomial identity $C(n,k) = C(n,n-k)$ on the result:
$$\kappa_j - \kappa_j^{\text{(dec)}}  = \frac{j!}{\beta_s^j} \sum_{k=2}^j \frac{\kappa_k(T_m)}{k!} \binom{j-1}{k-1} \lambda_s^k$$
The relative error for $j>1$ is
$$\frac{\kappa_j - \kappa_j^{\text{(dec)}}}{\kappa_j^{\text{(dec)}}} = \frac{1}{\expect{T_m}}\sum_{k=2}^j \frac{\kappa_k(T_m)}{k!} \binom{j-1}{k-1} \lambda_s^{k-1}$$
Contributions from $k \ge 3$ are negligible in the regime of interest, as they are doubly suppressed: they involve higher powers of $\lambda_s$, which must be small for the slow mechanism to remain infrequent, and higher-order cumulants $\kappa_k(T_m)$ ($k \ge 3$), which are negligible for the quasi-Gaussian $T_m$ distribution (Sec.~\ref{sec:Normal Approximation}). 
Then
\begin{equation}\label{eq:errorCumulantPure}
	\frac{\kappa_j - \kappa_j^{\text{(dec)}}}{\kappa_j^{\text{(dec)}}} \approx (j-1)\,\frac{\lambda_s \Var{T_m}}{2\expect{T_m}} 
\end{equation}
For the variance ($j=2$), this expression is identical to the error ratio that can readily be obtained by applying the law of total variance, which confirms the consistency of this more general cumulant-based approach.

Substituting in Eq. \eqref{eq:errorCumulantPure} the expressions for $\expect{T_m}$ (Eq. \eqref{eq: ExpectTm}), $\Var{T_m}$ (Eq. \eqref{eq: VarTm}), and  the mean number of slow-retention events per particle $\Lambda_s := \lambda_s\,\expect{T_m}$:
\begin{equation}
	\frac{\kappa_j - \kappa_j^{\text{(dec)}}}{\kappa_j^{\text{(dec)}}}
	\approx 
	(j-1)\,\Lambda_s\left[\frac{1}{\mathrm{Pe}} + \frac{1}{2}\left(\frac{\sigma_0}{x}\right)^2\right]
\end{equation}
The numerical value of the dimensionless bracketed quantity, under standard chromatographic conditions (high Péclet number and $\sigma_0/x \ll 1$), is much smaller than unity. In addition, the slow mechanism is modeled as infrequent at the single-particle level, so that $\Lambda_s$ remains small enough to generate tailing. Hence, for any fixed low order $j$, the relative error in the pure cumulants remains small throughout the parameter regime in which the model is intended to operate.

The preceding analysis is limited to the pure cumulants of $T_{s,s}$, whereas those of the total elution time $T_G + T_{s,s}$ additionally depend on mixed cumulants $\kappa_{p,q}(T_G,T_{s,s})$ for ($p,q \ge 1$). These arise from the shared dependence of $T_G$ and $T_{s,s}$ on $T_m$ in the exact model, but vanish in the decoupled case ($T_G + T'_{s,s}$).   It is therefore useful to assess their relative magnitude in order to quantify the impact of the decoupling approximation on the cumulants of the total elution time.

The joint CGF is defined as $K_{G,s}(u_G,u_s) := \log \expect{e^{u_G T_G + u_s T_{s,s}}}$. Conditioned on $T_m$, the independence between $T_G$ and $T_{s,s}$ implies
\[
K_{G,s}(u_G,u_s)
= \log \mathbb{E}_{T_m}\left[
e^{K_{T_G \mid T_m}(u_G) + K_{T_{s,s} \mid T_m}(u_s)}
\right].
\]
The conditional CGFs are $K_{T_{s,s} \mid T_m}(u_s) = T_m c(u_s)$ and
\[
K_{T_G \mid T_m}(u_G) = T_m\,p(u_G),
\quad
p(u_G) := (1+\mu_f)\,u_G + \frac{\sigma_f^2}{2}\,u_G^2,
\]
which is linearly proportional to $T_m$ because it is the sum of the CGF of the $T_m$ term itself ($T_m u_G$) and the $T_{s,f}$ CGF, whose conditional parameters $\mu_f T_m$ and $\sigma_f^2 T_m$ scale with $T_m$, as established in Sec. 4.2.

Substitution yields
\begin{align}
	K_{G,s}(u_G,u_s)
	&= \log \mathbb{E}_{T_m}\left[
	\exp\!\big( T_m [p(u_G) + c(u_s)] \big)
	\right] \\
	&= K_{T_m}\!\big(p(u_G) + c(u_s)\big),
\end{align}
whose expansion in the cumulants of $T_m$ is
\[
K_{G,s}(u_G,u_s)
= \sum_{k=1}^\infty \frac{\kappa_k(T_m)}{k!}\,[p(u_G) + c(u_s)]^{k}.
\]

The $k=1$ contribution, $\kappa_1(T_m)\,[p(u_G) + c(u_s)]$,
is additively separable in $u_G$ and $u_s$ and consequently yields no mixed cumulants; all $\kappa_{p,q}(T_G,T_{s,s})$ with $p$ and $q$ positive originate from the $k \ge 2$ terms. The main contribution corresponds to the $k=2$ term that can be expressed as:
\[ \frac{\Var{T_m}}{2}\,\big[p(u_G)^2 + 2 p(u_G)c(u_s) + c(u_s)^2\big].\]
The leading mixed cumulants are associated with the cross term $\Var{T_m}\,p(u_G)c(u_s)$, that can be expanded as
\[
p(u_G)c(u_s)
=\lambda_s \sum_{m=1}^{\infty}\frac{u_s^{m}}{\beta_s^{m}}
\Big[(1+\mu_f)\,u_G + \tfrac{\sigma_f^2}{2}\,u_G^2\Big].
\]
At this level of approximation, the only non–vanishing mixed cumulants are those with $p=1$ or $p=2$:
\begin{align}
	\kappa_{1,q}(T_G,T_{s,s})&= (1+\mu_f)\,\lambda_s\,\Var{T_m}\,\frac{q!}{\beta_s^{q}},
	\quad &q\ge 1,\\	
	\kappa_{2,q}(T_G,T_{s,s})
	&= \sigma_f^{2}\,\lambda_s\,\Var{T_m}\,\frac{q!}{\beta_s^{q}},
	\quad &q\ge 1.
\end{align}
Mixed cumulants with $p\ge 3$ vanish at $k=2$ because $p(u_G)$ is at most quadratic in $u_G$.

To assess the impact of these mixed cumulants, $\kappa_{p,q}(T_G,T_{s,s})$, their magnitude is compared against the pure slow-mechanism contribution of the same total order $j=p+q$, $\kappa_j^{\text{(dec)}}(T'_{s,s})$:
\begin{align}
	\frac{\kappa_{1,q}(T_G,T_{s,s})}{\kappa_{q+1}^{\text{(dec)}}(T'_{s,s})}
	&= (1+\mu_f)\,\frac{\Var{T_m}}{\expect{T_m}}\,
	\frac{\beta_s}{q+1},
	\\[4pt]
	\frac{\kappa_{2,q}(T_G,T_{s,s})}{\kappa_{q+2}^{\text{(dec)}}(T'_{s,s})}
	&= \sigma_f^{2}\,\frac{\Var{T_m}}{\expect{T_m}}\,
	\frac{\beta_s^{2}}{(q+1)(q+2)}.
\end{align}
Provided that the fast–mechanism parameters remain in standard chromatographic ranges, these expressions show that each mixed cumulant is only a small fraction of the corresponding slow–mechanism cumulant in the decoupled model. In this regime, neglecting these mixed cumulants, an implicit result of the decoupling approximation, has a quantitatively minor impact on the cumulants of the total elution time.

\subsection{Final Composite Density}
The PDF of the elution time, $T$, modeled as the sum of two independent variables, $T_G$ and $T_{s,s}'$, is given by their convolution:
\begin{equation}\label{eq:final_expresion}
f_T(t) \approx e^{-\Lambda_s}\sum_{\ell=0}^\infty \frac{\Lambda_s^{\,\ell}}{\ell!}\, h_\ell(t)
\end{equation}
where $h_0(t):=\mathcal{N}(t;\mu_G,\sigma_G^2)$ and, for $\ell \ge 1$, $h_\ell(t)$ is the Normal-Gamma convolution. A closed-form expression in terms of the confluent hypergeometric function of the first kind, ${}_1\mathcal{F}_1(a;b;z)$, is given by:
\begin{widetext}
\begin{equation}
h_\ell(t) = 2^{-\ell/2}\beta_s^\ell\sigma_G^{\ell-2} e^{-\frac{(t-\mu_G)^2}{2\sigma_G^2}} \left[ \frac{\sigma_G}{\sqrt{2}\,\Gamma\left((\ell+1)/2\right)}\, {}_1\mathcal{F}_1\left(\frac{\ell}{2};\frac12;Z\right)+ \frac{t-\mu_G-\beta_s\sigma_G^2}{\Gamma(\ell/2)}\,{}_1\mathcal{F}_1\left(\frac{\ell+1}{2};\frac32;Z\right) \right]
\end{equation}
\end{widetext}
\noindent with 
\[Z := \frac{(t-\mu_G-\beta_s\sigma_G^2)^2}{2\sigma_G^2}. \] 
This expression can be evaluated with standard scientific computing libraries. As an implementation note, for typical HPLC conditions, large arguments can be encountered and arbitrary-precision arithmetic may otherwise be required. 
To avoid this and to significantly reduce computational cost, Kummer’s transformation, $\exp(z)\,{}_1\mathcal{F}_1(a;b;-z)={}_1\mathcal{F}_1(b-a;b;z)$, can be applied. This is particularly useful when numerous evaluations are needed (e.g., peak-parameter estimation). An implementation of this expression can be found in ref. \citenum{STEPcode}.

\subsection{Validation of the Functional Form against Experimental Data}
To assess the appropriateness of Eq. \eqref{eq:final_expresion} for describing experimental peak shapes, two illustrative cases were examined. Moderately asymmetric peaks were selected for analysis, as the proposed form reduces to a Gaussian for symmetric profiles and thus offers no specific advantage in those cases. 
One is a cyclohexane peak from gas chromatography, presented in Fig. 6 of Ref. \citenum{pap2001}. The other corresponds to the fluoride peak obtained via ion chromatography, shown in Application Note \cite{thermoAN003548}.
Numerical values were obtained by digitizing published figures using the software Engauge Digitizer\cite{mark_mitchell_2020_3941227}, and analyses were confined to the peak region delimited by 1\% relative-height thresholds.
Model parameters for the proposed function, and for the established EMG\cite{Kalambet2011} used as reference, were estimated via nonlinear least squares fitting implemented using the SciPy\cite{2020SciPy-NMeth} library.
The appropriateness of the function was assessed through visual inspection of the fitted curves and residual plots (Fig. \ref{fig:fits}), complemented by quantitative comparison of the RSE.

\begin{widetext}
	\centering
	\includegraphics[width=1.\linewidth]{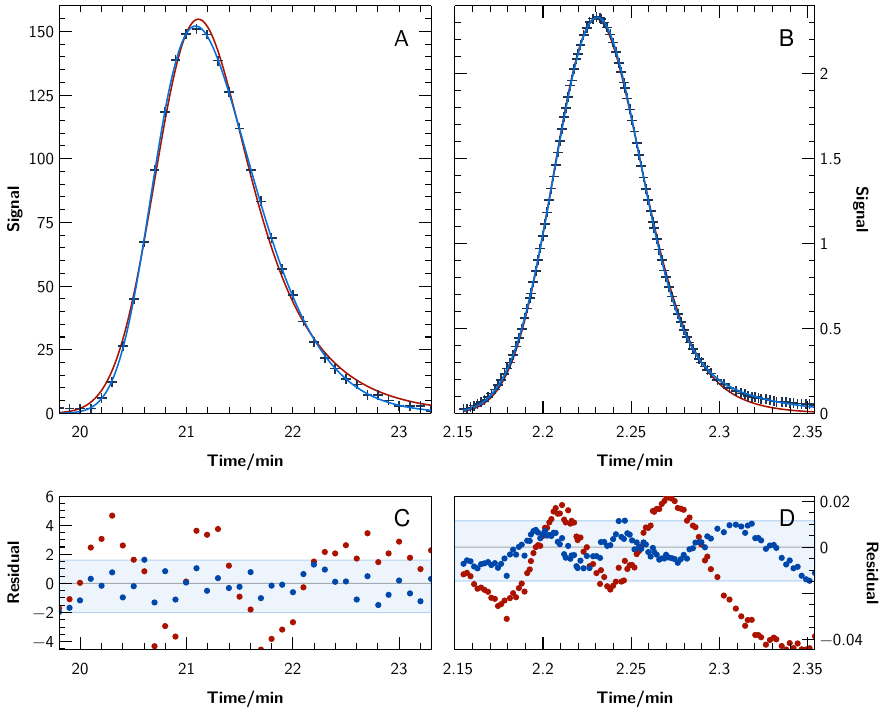}
	\captionof{figure}{Graphical representation of least-squares fits obtained using the proposed function (blue line) and the EMG (red line) applied to experimental data (gray crosses). Panels A and C correspond to a  cyclohexane peak from gas chromatography (see Ref. \citenum{pap2001}); Panels B and D correspond to a fluoride peak from ion chromatography (see Ref. \citenum{thermoAN003548}). Top panels (A–B): fitted curves. Bottom panels (C–D): residuals.}
	\label{fig:fits}
\end{widetext}
Substantially smaller RSE values were obtained for the proposed expression than for the EMG on the same data sets. Rapid convergence of the series in Eq. \eqref{eq:final_expresion} was observed: with a truncation level of $\ell_{\max}=3$, improvement factors of 2.72 (cyclohexane) and 3.52 (fluoride) were achieved. Increasing to $\ell_{\max}=6$ (the level used to compute the results in Fig. \ref{fig:fits}) only marginally refined these values--2.90 and 3.52, respectively--indicating that few terms suffice for convergence, allowing for a computationally inexpensive evaluation.
The residuals (Fig. \ref{fig:fits}, panels C–D) for the EMG are positive near the leading/trailing regions in cyclohexane, negative in fluoride; this opposite behaviour is not seen with the proposed form.
For a broader comparison, additional fits were performed using the PeakFit software against all its implemented chromatographic peak-shape functions (Gaussian, HVL, NLC, Giddings, EMG, GMG, EMG+GMG, GEMG-4, GEMG-5, log-normal-4, and the fronted/tailed Eval4 forms). RSE were normalised by the corresponding peak height. 
For the fluoride peak, the smallest normalised RSE (0.26\%) was obtained with the proposed expression, followed by EMG+GMG (0.47\%) and EMG (0.90\%), while all remaining functions lay in the 1.07–3.13\% range. For the cyclohexane peak, a normalised RSE of 0.64\% was achieved, slightly above the best empirical fits GMG, GEMG-5 and log-normal-4 (0.50–0.56\%), followed by GEMG-4 (0.86\%), with all other functions yielding larger errors between 1.44\% and 4.58\%.

\subsection{Tailing from Heterogeneous Slow Kinetics in a Comprehensive Transport Model}
In this section, the analysis of tailing from Sec.~\ref{sec: tailing} is extended to the previous more comprehensive model that incorporates axial diffusion and a fast-kinetic retention mechanism.
Herein, heterogeneity refers exclusively to the slow retention mechanism. A system will be termed heterogeneous if characterized by two distinct slow-kinetic mechanisms ($A$ and $B$), and homogeneous if characterized by a single such mechanism ($A$).
To ensure a meaningful comparison, both systems are modeled with an identical Gaussian component ($T_G$)---the mobile-phase first-passage time plus the fast-site contribution---with a common expected value $\mu_G$ and variance $\sigma_G^2$. Furthermore, in analogy with the treatment in Sec.~\ref{sec: tailing}, the total expected elution time  is preserved by enforcing the same expectation for the total slow-component contribution.
Due to the linearity of the expectation operator, the $T_G$ component cancels out in the mean-preservation constraint. Following a procedure analogous to that of Sec.~\ref{sec: tailing}, elementary algebraic manipulations yield the heterogeneous-to-homogeneous skewness ratio, $R$:
\begin{equation}
	R(\theta)\;=\;\omega_2\left(\frac{1+\theta}{\omega_1+\theta}\right)^{3/2}
\end{equation}
where $\omega_n := q_A + (1-q_A)/r^n$ for $n \in \{1,2\}$, and $\theta$ is the non-negative ratio of the variance $\sigma_G^2$ to the variance of the slow component in the homogeneous case,  $2\Lambda_A/\beta_A^2$. $R$ is always positive. In the absence of diffusion and fast sites ($\theta=0$) it correctly reduces to the base-model result. Its derivative can be expressed as
\begin{equation}
R^\prime(\theta)\;= \frac{3}{2}R(\theta)\;\frac{\omega_1-1}{(\omega_1+\theta)(1+\theta)}
\end{equation}
Provided $\omega_1$ and $\omega_2$, the sign of $R'(\theta)$ is constant in $\theta$ and equals $\mathrm{sign}(\omega_1-1)$. Hence $R$ is strictly decreasing in $\theta$ if $r>1$  (i.e., $\omega_1<1$), constant if $r=\omega_1=1$, and strictly increasing if $r<1$ ($\omega_1>1$). Moreover, setting $R(\theta)=1$  yields the unique threshold
\begin{equation}\label{eq:thresholdTailing}
\theta^{*}=\frac{\omega_2^{-2/3}\,\omega_1-1}{\,1-\omega_2^{-2/3}\,}
\end{equation}
such that $R(\theta)\le 1$ if $\theta\ge\theta^{*}$.

For $r<1$, the base case implies greater skewness in the heterogeneous system ($R(0)>1$) as shown in Sec. \ref{sec: tailing}. This  persists in the extended model, as in this case there is no positive $\theta^{*}$ satisfying Eq.~\eqref{eq:thresholdTailing}. Indeed, the condition $r<1$ implies $\omega_1, \omega_2 > 1$, ensuring the denominator in Eq.~\eqref{eq:thresholdTailing} is positive. A positive $\theta^{*}$ would thus require  $\omega_1^{3/2} > \omega_2$.
Let $\eta:=r^{-1}$, which is greater than $1$ in this case.
By Jensen’s inequality for the convex function $x\mapsto x^{3/2}$,
\begin{align*}
\omega_1^{3/2}&=\left[q_A+(1-q_A)\eta\right]^{3/2}\le q_A\cdot 1^{3/2}+(1-q_A)\eta^{3/2}\\
&< q_A+(1-q_A)\eta^{2}=\omega_2,
\end{align*}
which proves that such a $\theta^{*}$ does not exist.
Since $R$ is strictly increasing, the heterogeneous asymmetry increases further relative to the base case as $\sigma_G^{2}$ grows, with $R(\theta)>R(0)>1$ for all $\theta>0$ and $\lim_{\theta\to\infty}R(\theta)=\omega_2>1$.

For $r>1$, $R'(\theta)<0$, and thus $R(\theta)$ decreases monotonically.
Even if $R(0)>1$, there exists a unique $\theta^{*}>0$ beyond which $R\le 1$, thereby expanding the region in which heterogeneity reduces tailing.

These results imply that the contribution from diffusion and the fast-kinetic mechanism reinforces the base-case reduction of tailing.

\section{Conclusions}
A stochastic–diffusive framework for linear, isocratic and isothermal chromatography was developed. It connects microscopic retention events and advective–diffusive transport to an analytic time-domain expression representing the peak shape of a single analyte. 
In contrast to classical stochastic formulations built upon macroscopic kinetic rate equations, the present approach takes the probability laws of single-particle residence times as the primary objects. The peak profile is constructed stepwise from these laws--through the distribution of event counts and conditional residence-time distributions--so that each qualitative feature of the peak shape can be traced directly to specific probabilistic assumptions. This rigorous probabilistic construction provides a consistent basis for future theoretical extensions.

The foundational Poisson assumption for retention events was examined. The analysis yields an explicit expression for the excess variance, directly quantifying how the variance and temporal correlations of the microscopic rate fluctuations lead to deviations from the ideal Poisson model. This quantitative aspect has been overlooked in classical treatments.

The connection between macroscopic observables and the number of microscopic retention events was established by developing a statistical interpretation of the HETP. For non-pathological peaks under explicitly specified conventional operating conditions, the HETP can be interpreted as the column-segment length for which the variance of the residence time equals the square of its expectation.
Combining this interpretation with the compound Poisson model yields a conservative lower bound for the expected number of retention events. For exponentially distributed event durations, this bound simplifies to a direct function of only the experimentally accessible plate number and retention factor.

This bound is used to provide a theoretical justification, via a central-limit argument for random sums, for modeling the fast-kinetic retention contribution using a Gaussian function. 
The coupling of this contribution with advective–diffusive transport results in a mobile-plus-fast component which was identified as the PDF of a NIG distribution, whose convergence to a Gaussian surrogate under high-Péclet conditions was established through an analysis of its cumulant generating function.
This result, valid for idealized injections, was then generalized to incorporate finite initial spatial variance by analyzing its asymptotic limit.
For the infrequent slow-release mechanism, the contribution is incorporated by convolution with a Poisson–Gamma component after quantifying the decoupling-induced error in both pure and mixed cumulants, which are shown to be quantitatively negligible under typical chromatographic parameters.

The culmination of this development is an analytic, time-domain expression in terms of physically interpretable parameters to represent the peak shape for a single analyte under isocratic and isothermal conditions.  
Through this expression, significant limitations of Laplace-domain models are overcome.
In the two illustrative experimental peaks considered, fitting with this expression yielded the lowest average normalised RSE among the twelve chromatographic peak-shape functions implemented in the PeakFit software, lending additional support to the underlying theoretical approach and to the use of a mechanistically grounded peak-shape description.

The influence of mechanistic heterogeneity on peak asymmetry was also analyzed. It was demonstrated that when kinetic retention effects are isolated from confounding broadening mechanisms at a fixed mean retention time, heterogeneity does not inherently increase tailing; indeed, under specific and physically relevant conditions, it can lead to a reduction in skewness. The analysis was extended to the more complete advection-diffusion model, wherein the parameter region for heterogeneity-induced tailing reduction was shown to be further expanded by the Gaussian contribution from transport and fast-site kinetics.

\section*{Acknowledgement}
This work was supported by the National Scientific and Technical Research Council of Argentina (CONICET).
\linespread{1.05} 

\printbibliography
\end{multicols} 
\end{document}